\documentclass[aps,prl,showpacs,twocolumn,superscriptaddress]{revtex4}
\usepackage{bm,color}
\usepackage{amsmath}
\usepackage{braket}
\usepackage{color}
\usepackage[pdftex]{graphicx}
\usepackage{comment}
\begin{document}
\title{Predicted Photo-Induced Topological Phases in Organic Salt $\alpha$-(BEDT-TTF)$_2$I$_3$}	
\author{Keisuke Kitayama}
\affiliation{Department of Applied Physics, Waseda University, Okubo, Shinjuku-ku, Tokyo 169-8555, Japan}
\affiliation{Department of Physics, University of Tokyo, Tokyo 113-8656, Japan}
\author{Masahito Mochizuki}
\affiliation{Department of Applied Physics, Waseda University, Okubo, Shinjuku-ku, Tokyo 169-8555, Japan}
\begin{abstract}
The emergence of photo-induced topological phases and their phase transitions are theoretically predicted in organic salt $\alpha$-(BEDT-TTF)$_2$I$_3$, which possesses inclined Dirac cones in its band structure. By analyzing a photo-driven tight-binding model describing conduction electrons in the BEDT-TTF layer using the Floquet theorem, we demonstrate that irradiation with circularly polarized light opens a gap at the Dirac points, and the system eventually becomes a Chern insulator characterized by a quantized topological invariant. A rich phase diagram is obtained in plane of amplitude and frequency of light, which contains Chern insulator, semimetal, and normal insulator phases. We find that the photo-induced Hall conductivity provides a sensitive means to detect the predicted phase evolutions experimentally. This work contributes towards developing the optical manipulation of electronic states in matter through broadening the range of target materials that manifest photo-induced topological phase transitions.
\end{abstract}
\maketitle
%\sloppy 

\section{Introduction}
%%%%%%%%%%%%%%%%%%%%%%%%%%%%%%%%
\begin{figure}[htb]
\includegraphics[scale=0.5]{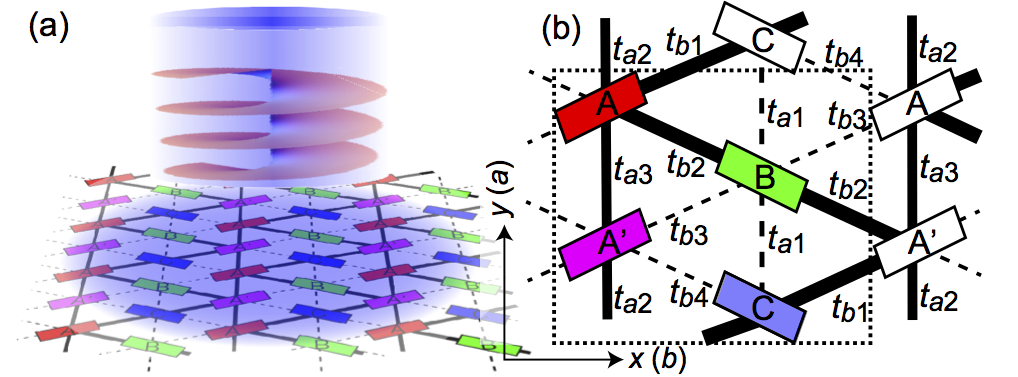}
\caption{(a) Schematics of $\alpha$-(BEDT-TTF)$_2$I$_3$ irradiated with circularly polarized light. (b) Unit cell (dashed rectangle) with four nonequivalent BEDT-TTF molecules (A, A$^\prime$, B, C) and transfer integrals in the BEDT-TTF layer.}
\label{Fig1}
\end{figure}
%%%%%%%%%%%%%%%%%%%%%%%%%%%%%%%%
Photo-induced phase transitions are one of the central topics in recent condensed-matter physics~\cite{Tokura06,Bukov15,Basov17,Aoki14,Yonemitsu06}. A theoretical study using the Floquet theorem has predicted that the honeycomb lattice irradiated with circularly polarized light attains a topological band structure similar to the band structure originally proposed by Haldane~\cite{Haldane88} that exhibits a topological phase transition. This transition results in the photo-induced quantum Hall effect in graphene, even in the absence of an external magnetic field~\cite{Oka09,Kitagawa11} and was indeed confirmed in a recent experiment~\cite{McIver19}. Since this pioneering theoretical work, the Floquet theory has been applied to various electron systems and has revealed many interesting photo-induced topological phenomena~\cite{Kitagawa10,Lindner11,Grushin14,ZhengW14,Mikami16,JYZou16,Ezawa13,Kang20,Claassen16,ZhangMY19, ZYan16,Sato16,Kitamura17,LDu17,Ezawa17,Takasan15,Takasan17a,Takasan17b,Menon18,ChenR18}. The topic is now attracting enormous research interest from the viewpoint of both fundamental science and electronics applications.

However, most of the previous research has dealt with tight-binding models on simple lattices such as the honeycomb lattice~\cite{Oka09,Kitagawa11,Kitagawa10}, the Kagome lattice~\cite{Mikami16,LDu17}, the Lieb lattice~\cite{Mikami16}, and the stacked graphene systems~\cite{JYZou16} or simple materials with a two-dimensional atomic layer such as graphene~\cite{Oka09,Kitagawa11,Kitagawa10}, silicene~\cite{Ezawa13}, black phosphorene~\cite{Kang20} and transition-metal dichalcogenides~\cite{Claassen16,ZhangMY19}. There have been few studies based on realistic models for specific materials. In addition, the only successful experiment so far was done for graphene~\cite{McIver19}, which has a simple electronic structure described well by the tight-binding model on honeycomb lattice~\cite{Haldane88,CastroNeto09}. However, to develop further this promising research field, widening the range of target materials is indispensable, and towards this objective, theoretical studies on actual materials with complex electronic and crystalline structures are highly desired. Moreover, we can expect richer material-specific photo-induced topological phenomena in studies on actual materials. One promising material is an organic salt $\alpha$-(BEDT-TTF)$_2$I$_3$ where BEDT-TTF denotes bis(ethylenedithio)-tetrathiafulvalene~\cite{Tajima06}. This compound has inclined Dirac cones in its band structure~\cite{Katayama06,Kobayashi07,Kajita14}, and many interesting topological properties and phenomena rising from these Dirac-cone bands have been theoretically investigated so far, e.g., the quantum Hall effect~\cite{Kajita14}, the structures of Berry curvature in momentum space~\cite{Suzumura11}, and the flux-induced Chern insulator phases~\cite{Osada17}.

In this paper, we theoretically predict the emergence of photo-induced topological phases and their phase transitions in $\alpha$-(BEDT-TTF)$_2$I$_3$. By applying the Floquet theory to the photo-driven tight-binding model for conduction electrons in the BEDT-TTF layer, we demonstrate that the inclined Dirac cones become gapped at the Dirac points by the irradiation with circularly polarized light [see Fig.~\ref{Fig1}(a)]. The system then becomes a topological insulator characterized by a quantized Chern number~\cite{Thouless82,Kohmoto85} and conductive chiral edge states~\cite{Hao08}. We obtain a rich phase diagram in the plane of the amplitude and frequency of the light that contains phases corresponding to a Chern insulator, semimetal, and normal insulator. The calculated photo-induced Hall conductivity shows characteristic dependencies on the light amplitude and temperature in each phase, indicating that this quantity provides a sensitive experimental indicator in the detection and identification of these topological phases and their phase transitions. One advantage of the usage of organic compounds is that an effective amplitude of light is an order of magnitude larger than graphene because their lattice constants are much larger (to be discussed later). They enhance the feasibility of the experiments. This work contributes to the development of this field by expanding the potential range of materials for research.

\section{Model and Method}
We employ a tight-binding model to describe the electronic structure of $\alpha$-(BEDT-TTF)$_2$I$_3$~\cite{Katayama06,Kajita14}, which is given by,
%%%%%%%%%%%%%%%%%%%%%%%%%%%%
\begin{eqnarray}
H=\sum_{i,j}\sum_{\alpha, \beta} t_{i\alpha,j\beta} c^\dagger_{i\alpha}c_{j\beta}.
\end{eqnarray}
%%%%%%%%%%%%%%%%%%%%%%%%%%%%
The unit cell of the BEDT-TTF layer contains four molecular sites (A, A$^\prime$, B, C), and we consider transfer integrals $t_{i\alpha,j\beta}$ among them [Fig.~\ref{Fig1}(b)] where $i$ and $j$ label the unit cells, whereas $\alpha$ and $\beta$ label the molecular sites.  Under ambient pressure, this compound exhibits a charge-ordered ground state~\cite{Tajima06}. When a uniaxial pressure is applied along the $a$ axis, this charge order melts and eventually inclined Dirac cones emerges within its peculiar band structure. In this study, we consider the latter situation by taking the transfer integrals evaluated theoretically at uniaxial pressure $P_{a}$ of 4 kbar; specifically,
$t_{a1}=-0.038$ eV, 
$t_{a2}= 0.080$ eV, 
$t_{a3}=-0.018$ eV, 
$t_{b1}= 0.123$ eV, 
$t_{b2}= 0.146$ eV, 
$t_{b3}=-0.070$ eV, and 
$t_{b4}=-0.025$ eV~\cite{Kobayashi04}.

After the Fourier transformations with respect to the spatial coordinates,
the tight-binding Hamiltonian is rewritten in the momentum space as,
%%%%%%%%%%%%%%%%%%%%%%%%%%%%
\begin{eqnarray}
H=\sum_{\bm{k}}(c^{\dagger}_{\bm{k}A}\, c^{\dagger}_{\bm{k}A'}\,c^{\dagger}_{\bm{k}B}\,c^{\dagger}_{\bm{k}C})\hat{H}(\bm{k})\left(
\begin{array}{c}
c_{\bm{k}A} \\
c_{\bm{k}A'} \\
c_{\bm{k}B} \\
c_{\bm{k}C}
\end{array}
\right)
\end{eqnarray}
%%%%%%%%%%%%%%%%%%%%%%%%%%%%
where
%%%%%%%%%%%%%%%%%%%%%%%%%%%%
\begin{eqnarray}
\hat{H}(\bm{k})=\left(
\begin{array}{cccc}
0 & A_2(\bm{k}) & B_2(\bm{k}) & B_1(\bm{k}) \\
A_2^{*}(\bm{k}) & 0 & B_2^{*}(\bm{k}) & B_1^{*}(\bm{k}) \\
B_2^{*}(\bm{k}) & B_2(\bm{k}) & 0 & A_1(\bm{k}) \\
B_1^{*}(\bm{k}) & B_1(\bm{k}) & A_1(\bm{k}) & 0
\end{array}
\right)
\label{alpham}
\end{eqnarray}
%%%%%%%%%%%%%%%%%%%%%%%%%%%%
with
%%%%%%%%%%%%%%%%%%%%%%%%%%%%
\begin{align}
&A_1(\bm{k})=2t_{a1}(\bm{k})\cos(k_y/2) \\
&A_2(\bm{k})=t_{a2}(\bm{k})e^{ik_y/2}+t_{a3}(\bm{k})e^{-ik_y/2} \\
&B_1(\bm{k})=t_{b1}(\bm{k})e^{i(k_x/2+k_y/4)}+t_{b4}(\bm{k})e^{-i(k_x/2-k_y/4)} \\
&B_2(\bm{k})=t_{b2}(\bm{k})e^{i(k_x/2-k_y/4)}+t_{b3}(\bm{k})e^{-i(k_x/2+k_y/4)}.
\end{align}
%%%%%%%%%%%%%%%%%%%%%%%%%%%%

We then consider a situation, in which this compound is irradiated with circularly polarized light applied perpendicular to the $ab$ plane [Fig.~\ref{Fig1}(a)]. The applied circularly polarized light generates a vector potential $\bm A(\tau)=A(\cos\omega\tau, \sin\omega\tau)$, which corresponds to a time-dependent electric field:
%%%%%%%%%%%%%%%%%%%%%%%%%%%%
\begin{eqnarray}
\bm E(\tau)=-\frac{d \bm A}{d \tau}=A\omega(\sin\omega\tau, -\cos\omega\tau).
\end{eqnarray}
%%%%%%%%%%%%%%%%%%%%%%%%%%%%
In the presence of this vector potential, the transfer integrals attain Peierls phases as,
%%%%%%%%%%%%%%%%%%%%%%%%%%%%
\begin{align}
&t_{i\alpha,j\beta}(\tau)
\nonumber \\
&=t_{i\alpha,j\beta}\exp{
\left[-i\frac{e}{\hbar}\bm{A}(\tau)\cdot (\bm{r}_{i\alpha}-\bm{r}_{j\beta})\right]}
\nonumber \\
&=t_{i\alpha,j\beta}\exp\left[i\left\{\mathcal{A}_b(\tilde{x}_j-\tilde{x}_i)\cos\omega\tau
+\mathcal{A}_a(\tilde{y}_j-\tilde{y}_i)\sin\omega\tau\right\}\right]
\end{align}
%%%%%%%%%%%%%%%%%%%%%%%%%%%%
Here we introduce dimensionless quantities $\mathcal{A}_a=eAa/\hbar$ and $\mathcal{A}_b=eAb/\hbar$ and dimensionless coordinates $\bm r_{i\alpha}=(b\tilde{x}_{i\alpha}, a\tilde{y}_{i\alpha})$ with $a$ and $b$ being the lattice constants along the $y$ and $x$ axes, respectively [Fig.~\ref{Fig1}(b)]. We use experimental values $a$=0.9187 nm and $b$=1.0793 nm~\cite{Mori12}, which give a ratio $\mathcal{A}_b/\mathcal{A}_a=b/a=1.175$. The amplitude of the ac electric field of light $E^\omega$ is given by $E^\omega=A\omega=\mathcal{A}_a\hbar\omega/ea$.

In the momentum representation, the effects of Peierls phases are taken into account by replacing the momenta $k_x$ and $k_y$ with $k_x+\mathcal{A}_b$ and $k_y+\mathcal{A}_a$, respectively. Then we obtain the time-dependent Hamiltonian for the photo-irradiated $\alpha$-(BEDT-TTF)$_2$I$_3$ as,
%%%%%%%%%%%%%%%%%%%%%%%%%%%%
\begin{eqnarray}
\hat{H}(\bm{k},\tau)=\left(
\begin{array}{cccc}
0 & A_2(\bm{k},\tau) & B_2(\bm{k},\tau) & B_1(\bm{k},\tau) \\
A_2^{*}(\bm{k},\tau) & 0 & B_2^{*}(\bm{k},\tau) & B_1^{*}(\bm{k},\tau) \\
B_2^{*}(\bm{k},\tau) & B_2(\bm{k},\tau) & 0 & A_1(\bm{k},\tau) \\
B_1^{*}(\bm{k},\tau) & B_1(\bm{k},\tau) & A_1(\bm{k},\tau) & 0
\end{array}
\right)
\label{alpham}
\end{eqnarray}
%%%%%%%%%%%%%%%%%%%%%%%%%%%%
with
%%%%%%%%%%%%%%%%%%%%%%%%%%%%
\begin{eqnarray}
A_1(\bm{k},\tau) &=& 2t_{a1}\cos\left(\frac{k_y}{2}+\frac{\mathcal{A}_a}{2}\sin\omega\tau \right) 
\\
A_2(\bm{k},\tau) &=&
t_{a2}\exp\left[ i\left(\frac{k_y}{2}+\frac{\mathcal{A}_a}{2}\sin\omega\tau \right)\right]
\nonumber \\ &+&
t_{a3}\exp\left[-i\left(\frac{k_y}{2}+\frac{\mathcal{A}_a}{2}\sin\omega\tau \right)\right]
\\
B_1(\bm{k},\tau) &=&
 t_{b1}\exp\left[ i\left(\frac{k_x}{2}+\frac{k_y}{4}\right)\right]
\exp\left[ i\mathcal{A}\sin(\omega\tau + \theta) \right]
\nonumber \\ &+&
t_{b4}\exp\left[-i\left(\frac{k_x}{2}-\frac{k_y}{4}\right)\right]
\exp\left[ i\mathcal{A}\sin(\omega\tau - \theta) \right]
\nonumber \\
\\
B_2(\bm{k},\tau) &=&
t_{b2}\exp\left[ i\left(\frac{k_x}{2}-\frac{k_y}{4}\right)\right]
\exp\left[-i\mathcal{A}\sin(\omega\tau - \theta) \right]
\nonumber \\ &+&
t_{b3}\exp\left[-i\left(\frac{k_x}{2}+\frac{k_y}{4}\right)\right]
\exp\left[-i\mathcal{A}\sin(\omega\tau + \theta) \right]
\nonumber \\
\end{eqnarray}
%%%%%%%%%%%%%%%%%%%%%%%%%%%%
where
%%%%%%%%%%%%%%%%%%%%%%%%%%%%
\begin{eqnarray}
\mathcal{A}=\frac{1}{4}\sqrt{4\mathcal{A}_b^2+\mathcal{A}_a^2},
\quad\quad
\theta=\tan^{-1}\left(\frac{2\mathcal{A}_b}{\mathcal{A}_a}\right).
\end{eqnarray}
%%%%%%%%%%%%%%%%%%%%%%%%%%%%

We analyze this time-periodic tight-binding Hamiltonian using the Floquet theory. The time-dependent Schr\"odinger equation is given by,
%%%%%%%%%%%%%%%%%%%%%%%%%%%%
\begin{eqnarray}
i\hbar \frac{\partial}{\partial t}\ket{\Psi(\bm k,\tau)}
=H(\bm k,\tau)\ket{\Psi(\bm k,\tau)}.
\end{eqnarray}
%%%%%%%%%%%%%%%%%%%%%%%%%%%%
According to the Floquet theorem, the wave function $\ket{\Psi(\tau)}$ is written in the form 
%%%%%%%%%%%%%%%%%%%%%%%%%%%%
\begin{eqnarray}
\ket{\Psi(\bm k,\tau)}=e^{-i\varepsilon \tau/\hbar}\ket{\Phi(\bm k,\tau)}
\end{eqnarray}
%%%%%%%%%%%%%%%%%%%%%%%%%%%%
where $\ket{\Phi(\bm k,\tau)}=\ket{\Phi(\bm k,\tau+T)}$. Here $T(=2\pi/\omega$) is the temporal period of the ac light field, and $\varepsilon$ is the quasi-energy. This equation is rewritten in the form, 
%%%%%%%%%%%%%%%%%%%%%%%%%%%%
\begin{eqnarray}
\sum_m \hat{\mathcal{H}}_{n,m}(\bm k)\ket{\Phi_{\nu}^m(\bm k)}
= \varepsilon^n_{\nu}(\bm k) \ket{\Phi_{\nu}^n(\bm k)},
\label{H-Mw}
\end{eqnarray}
%%%%%%%%%%%%%%%%%%%%%%%%%%%%
with
%%%%%%%%%%%%%%%%%%%%%%%%%%%%
\begin{eqnarray}
\hat{\mathcal{H}}_{n,m}(\bm k) \equiv 
\hat{H}_{n-m}(\bm k)-m\omega\delta_{n,m}\hat{1}_{4 \times 4}
\label{H-Mw2}
\end{eqnarray}
%%%%%%%%%%%%%%%%%%%%%%%%%%%%
where $\nu$ labels the eigenstates, and $n$ corresponds to the number of photons. The four-component vector $\ket{\Phi_{\nu}^n(\bm k)}$ rerpesents the $\nu$th Floquet state ($\nu$=1,2,3,4) in the $n$-photon subspace. We introduce the following Fourier coefficients,
%%%%%%%%%%%%%%%%%%%%%%%%%%%%
\begin{eqnarray}
\ket{\Phi_{\nu}^n(\bm k)}&=&
\frac{1}{T}\int_0^T \ket{\Phi_{\nu}(\bm k,\tau)}e^{in\omega\tau} d\tau,\\
\hat{H}_n(\bm k)&=&\frac{1}{T}\int_0^T \hat{H}(\bm k,\tau)e^{in\omega\tau} d\tau
\nonumber \\
&=&\left(
\begin{array}{cccc}
0 & A_{2,n}(\bm k) & B_{2,n}(\bm k) & B_{1,n}(\bm k) \\
A_{2,-n}^{*}(\bm k) & 0 & B_{2,-n}^{*}(\bm k) & B_{1,-n}^{*}(\bm k) \\
B_{2,-n}^{*}(\bm k) & B_{2,n}(\bm k) & 0 & A_{1,n}(\bm k) \\
B_{1,-n}^{*}(\bm k) & B_{1,n}(\bm k) & A_{1,n}(\bm k) & 0
\end{array}
\right)
\nonumber \\
\end{eqnarray}
%%%%%%%%%%%%%%%%%%%%%%%%%%%%
with
%%%%%%%%%%%%%%%%%%%%%%%%%%%%
\begin{align}
&A_{1,n}(\bm k)=
t_{a1}\,e^{ik_y/2}J_{-n}(\mathcal{A}_a/2)+t_{a1}\,e^{-ik_y/2}J_{n}(\mathcal{A}_a/2)
\\
&A_{2,n}(\bm k)=
t_{a2}\,e^{ik_y/2}J_{-n}(\mathcal{A}_a/2)+t_{a3}\,e^{-ik_y/2}J_{n}(\mathcal{A}_a/2)
\\
&B_{1,n}(\bm k)=
t_{b1}\,e^{ i(k_x/2+k_y/4)}J_{-n}(\mathcal{A})e^{-in\theta}
\nonumber \\ & \quad\quad\quad\quad
+t_{b4}\,e^{-i(k_x/2-k_y/4)}J_{-n}(\mathcal{A})e^{+in\theta}
\\
&B_{2,n}(\bm k)=
t_{b2}\,e^{ i(k_x/2-k_y/4)}J_{n}(\mathcal{A})e^{+in\theta}
\nonumber \\ & \quad\quad\quad\quad
+t_{b3}\,e^{-i(k_x/2+k_y/4)}J_{n}(\mathcal{A})e^{-in\theta}
\end{align}
%%%%%%%%%%%%%%%%%%%%%%%%%%%%

We solve the eigenequation~(\ref{H-Mw}) by restricting the number of photons to $|m|\le M_{\rm max}$ ($M_{\rm max}$=16 throughout the present work). Consequently, the Floquet-Hamiltonian matrix $\hat{\mathcal{H}}(\bm k)$ is composed of $(2M_{\rm max}+1)\times(2M_{\rm max}+1)$ block matrices $\hat{\mathcal{H}}_{n,m}(\bm k) \equiv \hat{H}_{n-m}(\bm k)-m\omega\delta_{n,m}\hat{1}_{4 \times 4}$. The total size of $\hat{\mathcal{H}}(\bm k)$ is $(8M_{\rm max}+4)\times (8M_{\rm max}+4)$ (132$\times$132 in the present work) because the size of each block matrix is 4$\times$4. Note that as the frequency $\omega$ is reduced, a Floquet matrix of larger size $|m|$ is required, typically of order $W/\hbar\omega$, where $W$ is the band width. Having adopted $|m|\le 16$ for $\alpha$-(BEDT-TTF)$_2$I$_3$ with $W\sim0.8$ eV, the obtained results are sufficiently accurate for $\hbar\omega \gtrsim$ 0.05 eV.

The Chern number of the $\nu$th band $N_{\rm Ch}^\nu$ ($\nu$=1,2,3,4) is related to the Berry curvature $B_z^{n\nu}(\bm k)$,
%%%%%%%%%%%%%%%%%%%%%%%%%%%%%%%%%%
\begin{eqnarray}
N_{\rm Ch}^\nu=\frac{1}{2\pi}\int\int_{\rm BZ}\;B_z^{0\nu}(\bm k) dk_xdk_y,
\end{eqnarray}
%%%%%%%%%%%%%%%%%%%%%%%%%%%%%%%%%%
where the Berry curvature $B_z^{n\nu}(\bm k)$ of the $\nu$th $n$-photon Floquet band  at each $\bm k$ point is given by
%%%%%%%%%%%%%%%%%%%%%%%%%%%%%%%%%%
\begin{align}
&B_z^{n\nu}(\bm k)=
\nonumber \\
&i\sum_{(m,\mu)}\frac{
\bra{\Phi_{\nu}^n(\bm k)}\frac{\partial \mathcal{H}}{\partial k_x}\ket{\Phi_{\mu}^m(\bm k)}
\bra{\Phi_{\mu}^m(\bm k)}\frac{\partial \mathcal{H}}{\partial k_y}\ket{\Phi_{\nu}^n(\bm k)}
-{c.c.}}
{[\varepsilon^m_\mu(\bm k)-\varepsilon^n_\nu(\bm k)]^2}.
\end{align}
%%%%%%%%%%%%%%%%%%%%%%%%%%%%%%%%%%
Here $\mathcal{H}$ denotes the matrix of the Floquet Hamiltonian, and $\varepsilon^n_\nu(\bm k)$ and $\ket{\Phi_{\nu}^n(\bm k)}$ the eigenenergy and eigenvector of the equation~(\ref{H-Mw}) with $\nu=1,2,3,4$ and $|n|\le 16$. The summation is taken over $m$ and $\mu$ where $(m,\mu)\ne(n,\nu)$; ``$c.c.$" denotes the complex conjugate of the first term of the numerator. In this work, the Chern numbers are calculated using a numerical technique proposed by Fukui, Suzuki, and Hatsugai in Ref.~\cite{Fukui05}.

\section{Results}
%%%%%%%%%%%%%%%%%%%%%%%%%%%%%%%
\begin{figure}[tb]
\includegraphics[scale=0.5]{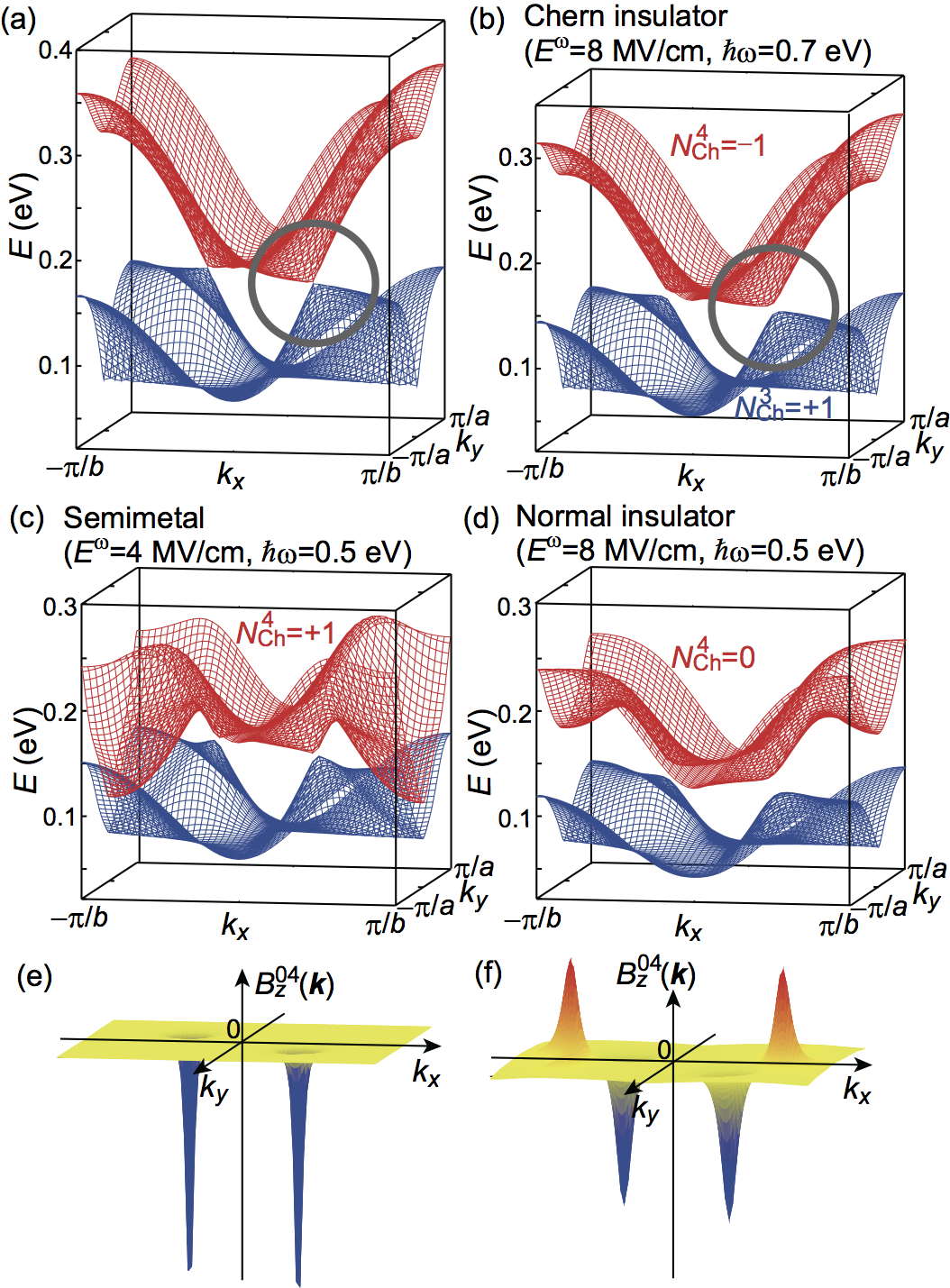}
\caption{(a) Band dispersions of the third and fourth bands, $E_3(\bm k)$ and $E_4(\bm k)$, before light irradiation. (b)-(d) Those for the photo-induced phases under irradiation with circularly polarized light, i.e., (b) Chern-insulator, (c) semimetal, and (d) normal-insulator phases. (e),(f) Berry curvatures of the fourth band $B_z^{04}(\bm k)$ in momentum space for (e) the photo-induced Chern insulator phase with $N_{\rm Ch}^4=-1$ and (f) the photo-induced normal insulator phase with $N_{\rm Ch}^4=0$.}
\label{Fig2}
\end{figure}
%%%%%%%%%%%%%%%%%%%%%%%%%%%%%%%%
We first discuss the photo-induced variation of band structures and their topological properties. Figure~\ref{Fig2}(a) shows the band dispersions for the third and fourth bands, $E_3(\bm k)$ and $E_4(\bm k)$, in the absence of photo-irradiation. These two bands make contact at two individual points in momentum space to form a pair of inclined Dirac cones. Note that the Dirac points are located at the Fermi level because this compound has a $3/4$ electron filling with three fully occupied lower bands and an unoccupied fourth band. Figures~\ref{Fig2}(b)-(d) show plots of $E_3(\bm k)$ and $E_4(\bm k)$ for photo-irradiated systems with various $E^\omega$ and $\omega$ of light. Once the system is irradiated with circularly polarized light, a gap opens at the Dirac points. These three band structures correspond to three different photo-induced phases characterized by the Chern number $N_{\rm Ch}$ and the band gap $E_{\rm gap}$, that is, (b) Chern-insulator, (c) semimetal, and (d) normal-insulator phases, respectively.

Here the band gap is defined by $E_{\rm gap}={\rm min}[E_4(\bm k)]-{\rm max}[E_3(\bm k)]$ where min[$E_4(\bm k)$] and max[$E_3(\bm k)$] are the minimum energy of the fourth band and the maximum energy of the third band, respectively. A positive $E_{\rm gap}$ means that the bulk is insulating, for which, in the whole momentum space, the fourth band is located above the Fermi level whereas the other three bands are located below the Fermi level. In contrast, a negative $E_{\rm gap}$ means that the bulk is semi-metallic, for which the third band is located above the Fermi level whereas the fourth band is located below the Fermi level in some portion of the momentum space. Importantly, when the electron filling $3/4$ with three fully occupied lower bands, a sum of the Chern numbers over three bands below the Fermi level, $N_{\rm Ch}=\sum_{\nu=1}^3 N_{\rm Ch}^\nu$, coincides with $-N_{\rm Ch}^4$ because conservation of invariance requires the sum of the Chern numbers over all four bands to be zero.

The band structure in Fig.~\ref{Fig2}(b) characterized by $E_{\rm gap}>0$ and $N_{\rm Ch}=-N_{\rm Ch}^4=+1$ is assigned to the Chern insulator phase. In contrast, the band structure in Fig.~\ref{Fig2}(c) is characterized by $E_{\rm gap}<0$, for which the fourth band around $\bm k=(\pm \pi,0)$ is lower in energy than the third band around the Dirac points. This band structure is assigned to the semimetal phase. Interestingly, the band structure in Fig.~\ref{Fig2}(d) has $E_{\rm gap}>0$ and resembles the band structure for the Chern insulator [Fig.~\ref{Fig2}(b)]. However, the Chern number $N_{\rm Ch}=-N_{\rm Ch}^4$ is zero in this case, indicating that the system lies in a topologically trivial insulating state. Therefore, this band structure is assigned to the normal insulator phase. Note that these phases appear as nonequilibrium steady states under the continuous application of circularly polarized light, and, in this sense, they are distinct from thermodynamically equilibrium phases.

We find a clear difference between the Chern insulator phase and the normal insulator phase in the Berry curvature $B_z^{04}(\bm k)$. The Berry curvature in the Chern insulator phase has two negative peaks around the gapped Dirac points [Fig.~\ref{Fig2}(e)], corresponding to a nonzero quantized Chern number $N_{\rm Ch}^4$ of $-1$, whereas that for the normal insulator phase has additional positive peaks as well as two negative peaks around the gapped Dirac points [Fig.~\ref{Fig2}(f)] that cancel the opposite contributions, resulting in $N_{\rm Ch}^4=0$. 

%%%%%%%%%%%%%%%%%%%%%%%%%%%%%%%%
\begin{figure}
\includegraphics[scale=0.5]{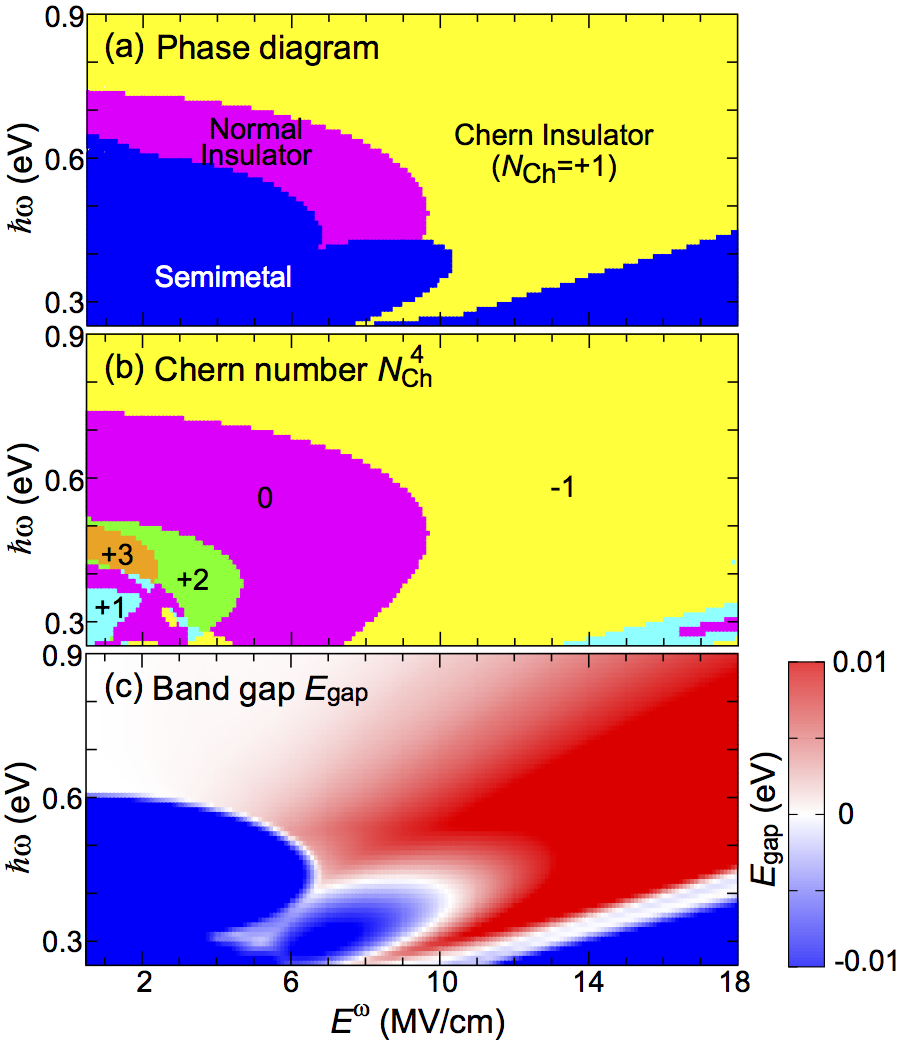}
\caption{(a) Phase diagram of photo-driven $\alpha$-(BEDT-TTF)$_2$I$_3$ in the plane of the amplitude $E^\omega$ and frequency $\omega$ of the applied circularly polarized light. (b), (c) Color maps of (b) the Chern number of the highest (fourth) band $N_{\rm Ch}^4$ and (c) the band gap $E_{\rm gap}$ in the plane of $E^\omega$ and $\omega$.}
\label{Fig3}
\end{figure}
%%%%%%%%%%%%%%%%%%%%%%%%%%%%%%%%
Figure~\ref{Fig3}(a) shows the phase diagram of photo-driven $\alpha$-(BEDT-TTF)$_2$I$_3$ in the plane of the amplitude $E^\omega$ and frequency $\omega$ of an applied circularly polarized light. Three phases are present, namely, the Chern-insulator, semimetal, and normal-insulator phases. This phase diagram is produced by calculating the Chern number of the (highest) fourth band $N_{\rm Ch}^4$ [Fig.~\ref{Fig3}(b)] and the band gap $E_{\rm gap}$ [Fig.~\ref{Fig3}(c)]. We assign the area with $E_{\rm gap}>0$ and $N_{\rm Ch}^4 \ne 0$ to the Chern insulator phase, the area with $E_{\rm gap}>0$ and $N_{\rm Ch}^4 = 0$ to the  normal insulator phase. The area with $E_{\rm gap}<0$ is assigned to the semimetal phase. According to the obtained phase diagram, we expect that the usage of high-frequency light with $\hbar\omega>0.75$ eV is preferable to observe the photo-induced Chern insulator phase in the low $E^\omega$ range, whereas the usage of low-frequency light with $\hbar\omega<0.7$ eV is suitable to observe the rich phase transitions upon the variation of $E^\omega$.

%%%%%%%%%%%%%%%%%%%%%%%%%%%%%%%%
\begin{figure}[tb]
\includegraphics[scale=0.5]{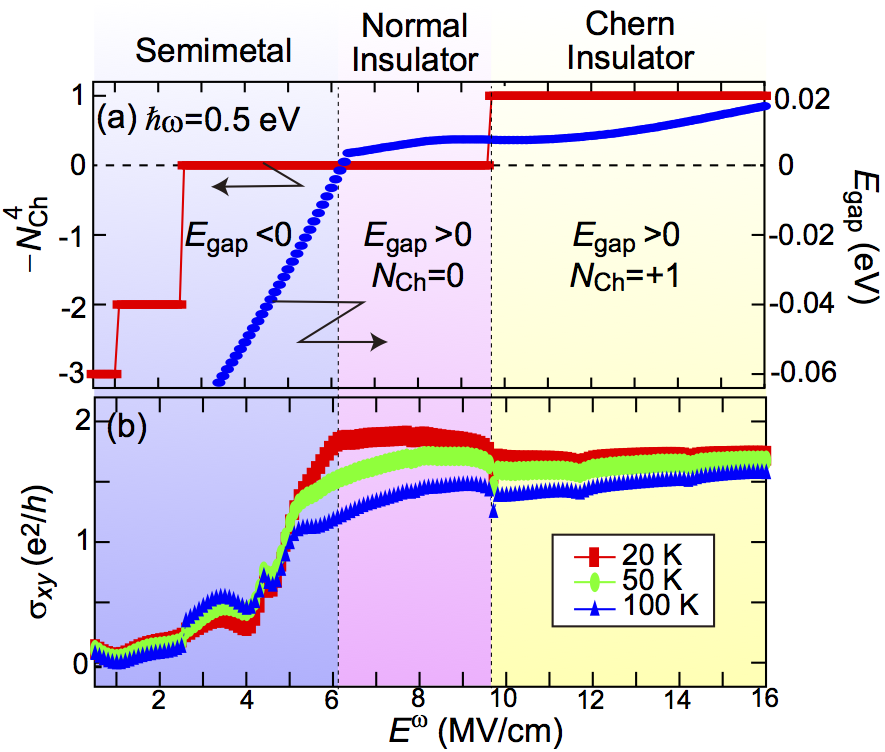}
\caption{(a) Chern number $-N_{\rm Ch}^4$ and band gap $E_{\rm gap}$ plotted as a function of the light amplitude $E^\omega$ when $\hbar\omega=0.5$ eV, which manifest successive emergence of three photo-induced electronic phases with increasing $E^\omega$. (b) $E^\omega$-profiles of the Hall conductivity $\sigma_{xy}$ for various temperatures when $\hbar\omega=0.5$ eV.}
\label{Fig4}
\end{figure}
%%%%%%%%%%%%%%%%%%%%%%%%%%%%%%%%
Finally, we discuss the Hall conductivity in the three different photo-induced electronic phases. This physical quantity is closely related to the topological nature of the electronic states and can be exploited to identify the predicted topological phases and to detect their phase transitions. We plot the calculated Chern number $-N_{\rm Ch}^4$ and band gap $E_{\rm gap}$ [Fig.~\ref{Fig4}(a)] as functions of the amplitude $E^\omega$ of the applied circularly polarized light setting $\hbar\omega=0.5$ eV, for which the three photo-induced phases, semimetal, normal-insulator, and Chern-insulator phases, emerge successively as $E^\omega$ increases. Recall that the relation $N_{\rm Ch}=-N_{\rm Ch}^4$ holds when the electron filling is $3/4$. We also plot the calculated $E^\omega$-profiles of the Hall conductivity $\sigma_{xy}$ for various temperatures when $\hbar\omega=0.5$ eV [Fig.~\ref{Fig4}(b)]. The Hall conductivity $\sigma_{xy}$ is calculated using the relation,
%%%%%%%%%%%%%%%%%%%%%%%%%%%%%%%%
\begin{eqnarray}
\sigma_{xy}=\frac{2e^2}{h}\int\int_{\rm BZ}\;\frac{dk_xdk_y}{2\pi}
\sum_{n,\nu} n_{n\nu}(\bm k) B_z^{n\nu}(\bm k)
\label{sgmyx}
\end{eqnarray}
%%%%%%%%%%%%%%%%%%%%%%%%%%%%%%%%
where the factor 2 accounts for the spin degeneracy. Here $n_{n\nu}(\bm k)$ is the nonequilibrium distribution function, which describes the electron occupations of Floquet bands in the photo-driven nonequilibrium steady states.

The nonequilibrium distribution function $n_{n\nu}(\bm k)$ is calculated using the Floquet-Keldysh formalism~\cite{Tsuji09,Aoki14}, which is formulated by combining the Keldysh Green's function technique~\cite{Jauho94,Mahan00} with the Floquet theory. The Dyson equation for the Green's function matrices is given by
%%%%%%%%%%%%%%%%%%%%%%%%%%%%%%%%%%%%%%%%%%%%%%%%%%%%%%%%%%%%%%%%%%%%
\begin{eqnarray}
& &\left(
\begin{array}{cc}
\hat{G}^{\rm R}(\bm k,\varepsilon) & \hat{G}^{\rm K}(\bm k,\varepsilon) \\
0                                    & \hat{G}^{\rm A}(\bm k,\varepsilon)
\end{array}
\right)^{-1}
\nonumber \\
& &=
\left(
\begin{array}{cc}
[\hat{G}^{\rm R0}(\bm k,\varepsilon)]^{-1} & 0 \\
0 & [\hat{G}^{\rm A0}(\bm k,\varepsilon)]^{-1}
\end{array}
\right)
-\left(
\begin{array}{cc}
\hat{\Sigma}^{\rm R} & \hat{\Sigma}^{\rm K}(\varepsilon) \\
0 & \hat{\Sigma}^{\rm A}
\end{array}
\right).
\nonumber \\
\end{eqnarray}
%%%%%%%%%%%%%%%%%%%%%%%%%%%%%%%%%%%%%%%%%%%%%%%%%%%%%%%%%%%%%%%%%%%%
Here $\hat{G}^{\rm R}$, $\hat{G}^{\rm A}$ and $\hat{G}^{\rm K}$ ($\hat{\Sigma}^{\rm R}$, $\hat{\Sigma}^{\rm A}$ and $\hat{\Sigma}^{\rm K}$) are matrices of the retarded, advanced, and Keldysh Green's functions (self-energies), respectively, each of which is composed of $(2M_{\rm max}+1)\times(2M_{\rm max}+1)$ block matrices where the size of each block matrix is 4$\times$4. The matrix components of $\hat{G}^{\rm R0}$, $\hat{G}^{\rm A0}$, $\hat{\Sigma}^{\rm R}$, $\hat{\Sigma}^{\rm A}$ and $\hat{\Sigma}^{\rm K}$ are given, respectively, by,
%%%%%%%%%%%%%%%%%%%%%%%%%%%%%%%%%%%%%%%%%%%%%%%%%%%%%%%%%%%%%%%%%%%%
\begin{align}
&[\hat{G}^{\rm R0}(\bm k,\varepsilon)]^{-1}_{n\nu,m\mu}
=\varepsilon\delta_{n,m}\delta_{\nu,\mu}-\mathcal{H}_{n\nu,m\mu}(\bm k)
\\
&[\hat{G}^{\rm A0}(\bm k,\varepsilon)]^{-1}_{n\nu,m\mu}
=\varepsilon\delta_{n,m}\delta_{\nu,\mu}-\mathcal{H}_{n\nu,m\mu}(\bm k)
\\
&[\hat{\Sigma}^{\rm R}]_{n\nu,m\mu}=-i\Gamma\delta_{n,m}\delta_{\nu,\mu}
\\
&[\hat{\Sigma}^{\rm A}]_{n\nu,m\mu}= i\Gamma\delta_{n,m}\delta_{\nu,\mu}
\\
&[\hat{\Sigma}^{\rm K}(\varepsilon)]_{n\nu,m\mu}=-2i\Gamma\tanh\left[
\frac{\varepsilon-\mu+m\omega}{2k_{\rm B}T} \right]\delta_{n,m}\delta_{\nu,\mu},
\end{align}
%%%%%%%%%%%%%%%%%%%%%%%%%%%%%%%%%%%%%%%%%%%%%%%%%%%%%%%%%%%%%%%%%%%%
where the symbol $\hat{M}_{n\nu,m\mu}$ denotes the $(\nu,\mu)$th component of the $(m,n)$th block matrix $\hat{M}_{nm}$ ($4 \times 4$), and the block matrix $\hat{\mathcal{H}}_{n,m}$ constituting the Floquet Hamiltonian is given by Eq.~(\ref{H-Mw2}). We consider a situation that the system is coupled to a heat reservoir at temperature $T$ with a dissipation coefficient $\Gamma$ where we set $\Gamma$=0.1 eV for the calculations. The lesser Green's function $\hat{G}^{<}$ and the lesser self-energy $\hat{\Sigma}^{<}$ are calculated, respectively, by,
%%%%%%%%%%%%%%%%%%%%%%%%%%%%%%%%%%%%%%%%%%%%%%%%%%%%%%%%%%%%%%%%%%%%
\begin{eqnarray}
& &\hat{G}^{<}(\bm k,\varepsilon)=\hat{G}^{\rm R}(\bm k,\varepsilon)
\;\hat{\Sigma}^{<}(\varepsilon)\;\hat{G}^{\rm A}(\bm k,\varepsilon),
\\
& &\hat{\Sigma}^{<}(\varepsilon)=
(\hat{\Sigma}^{\rm A}+\hat{\Sigma}^{\rm K}(\varepsilon)-\hat{\Sigma}^{\rm R})/2.
\end{eqnarray}
%%%%%%%%%%%%%%%%%%%%%%%%%%%%%%%%%%%%%%%%%%%%%%%%%%%%%%%%%%%%%%%%%%%%
The matrix components of the lesser self-energy $\hat{\Sigma}^{<}$ read,
%%%%%%%%%%%%%%%%%%%%%%%%%%%%%%%%%%%%%%%%%%%%%%%%%%%%%%%%%%%%%%%%%%%%
\begin{eqnarray}
[\hat{\Sigma}^{<}(\varepsilon)]_{n\nu,m\mu}=i\Gamma\left(1-\tanh\left[
\frac{\varepsilon-\mu+m\omega}{2k_{\rm B}T}
\right]\right)\delta_{n,m}\delta_{\nu,\mu}
\end{eqnarray}
%%%%%%%%%%%%%%%%%%%%%%%%%%%%%%%%%%%%%%%%%%%%%%%%%%%%%%%%%%%%%%%%%%%%
The nonequilibrium distribution function $n_{n\nu}(\bm k)$ for the $\nu$-th Floquet band in the $n$-photon subspace is given by,
%%%%%%%%%%%%%%%%%%%%%%%%%%%%%%%%%%%%%%%%%%%%%%%%%%%%%%%%%%%%%%%%%%%%
\begin{eqnarray}
n_{n\nu}(\bm k)=
\frac{\braket{\Phi_\nu^n(\bm k)|\hat{N}_{\bm k}(\varepsilon_\nu^n(\bm k))|\Phi_\nu^n(\bm k)}}
{\braket{\Phi_\nu^n(\bm k)|\hat{A}_{\bm k}(\varepsilon_\nu^n(\bm k))|\Phi_\nu^n(\bm k)}}
\end{eqnarray}
%%%%%%%%%%%%%%%%%%%%%%%%%%%%%%%%%%%%%%%%%%%%%%%%%%%%%%%%%%%%%%%%%%%%
where
%%%%%%%%%%%%%%%%%%%%%%%%%%%%%%%%%%%%%%%%%%%%%%%%%%%%%%%%%%%%%%%%%%%%
\begin{eqnarray}
& &\hat{A}_{\bm k}(\varepsilon)=
i(\hat{G}^{\rm R}(\bm k,\varepsilon)-\hat{G}^{\rm A}(\bm k,\varepsilon))/2\pi
\\
& &\hat{N}_{\bm k}(\varepsilon)=-i\hat{G}^{<}(\bm k,\varepsilon)/2\pi
\end{eqnarray}
%%%%%%%%%%%%%%%%%%%%%%%%%%%%%%%%%%%%%%%%%%%%%%%%%%%%%%%%%%%%%%%%%%%%
and $\ket{\Phi_\nu^n(\bm k)}$ is the Floquet eigenstates obtained by solving Eq.~(\ref{H-Mw}).

In Fig.~\ref{Fig4}(b), we find that the Hall conductivity $\sigma_{xy}$ is almost zero in the semimetal phase when $E^\omega$ is small ($E^\omega \lesssim 4$ MV/cm) but starts increasing with increasing $E^\omega$ at around $E^\omega \sim 4$ MV/cm towards the phase boundary to the normal insulator phase. It reaches a quantized value of $\sim 2e^2/h$ at the phase boundary and is kept constant in the normal insulator phase although it is slightly decreased as temperature increases. This nearly constant behavior does not change even when the system enters the Chern insulator phase, in which the value of $\sigma_{xy}$ is again nearly constant to be $\sim 2e^2/h$. The quantized Hall conductivity of $\sigma_{xy}\sim 2e^2/h$ in the Chern insulator phase is naturally understood from the quantized Chern number of $N_{\rm Ch}=-N_{\rm Ch}^4=1$. On the other hand, the finite $\sigma_{xy}$ in the nomarl insulator phase is rather nontrivial because the Chern number vanishes ($N_{\rm Ch}=0$) in this phase. It is attributable to the formation and the electron occupations of Floquet subbands (a series of replicas of the original bands with an energy spacing of $\hbar\omega$) in the present photo-driven system. Namely, when the light frequency is not large enough as in the present case with $\hbar\omega=0.5$ eV as compared to the bandwidth $W$ ($\sim$0.8 eV in the present $\alpha$-type organic salt), the hybridization and the anti-crossing of Floquet bands with different photon numbers occur. As a result, the highest Floquet band in the $m=-1$ subspace and the lowest Floquet band in the $m=+1$ subspace are partially occupied by electrons, and the Hall conductivity captures the Chern numbers of the Floquet bands in these finite-photon subspaces. This phenomenon cannot be expected in usual normal insulator phase in equilibrium.

Because both the normal insulator phase and the Chern insulator phase exhibit nearly the same behaviors of the Hall conductivity, it may be difficult to distinguish these phases by the Hall-conductivity measurements. However, the temperature dependence is more pronounced in the normal insulator phase. Hence, measurements of the precise temperature profiles of $\sigma_{xy}$ may be exploited for identification of the phases. It should be mentioned that in this $\alpha$-type organic salt, the Dirac points appear at low temperatures when the charge order melts. One possible option to realize such a situation is application of an uniaxial pressure $P_a$, which is known to melt the charge ordering in the ground state. Thus, the predicted topological phase transitions may be observed under application of the pressure. The measurements must be performed using diamond anvil pressure cells, which are transparent for laser light. Another option is to perform the experiments above the charge ordering temperature $T_{\rm CO}(\sim 40)$ K, which is much easier than the first option. Its feasibility is supported by our calculated Hall conductivity data, which shows that the quantized Hall conductivity is observed even at $T$=100 K.

\section{Conclusion}
To summarize, we theoretically predicted the emergence of rich photo-induced topological phases as nonequilibrium steady states in organic salt $\alpha$-(BEDT-TTF)$_2$I$_3$ with inclined Dirac-cone bands under continuous application of circularly polarized light. The predicted topological electronic phases and their transitions upon the laser-parameter variations are expected to be observed in future experiments by measuring the photo-induced Hall effect. A crucial advantage of the usage of organic compounds is that the effective amplitude of the laser is enhanced significantly due to the large molecule sizes and resulting larger lattice constants because the dimensionless laser amplitude $\mathcal{A}_a=eaE^\omega/\hbar\omega$ and $\mathcal{A}_b=ebE^\omega/\hbar\omega$ are proportional to the lattice constants $a$ and $b$. We expect the effective vector potential $\mathcal{A}$ in $\alpha$-(BEDT-TTF)$_2$I$_3$ to be nearly an order of magnitude larger than that for graphene, and hence the feasibility of experimental realization of the predicted photo-induced phenomena is expected to be increased. The field of photo-induced topological phase transitions is now rapidly growing, but target materials for research are still limited to a few toy models and simple atomic-layered materials. Our work may contribute by advancing this research field through broadening the range of target materials.

\section{Acknowledgment}
We thank Y. Tanaka for useful discussions. This work was supported by JSPS KAKENHI (Grant Nos. 17H02924, 16H06345, 19H00864, 19K21858, 19K23427, and 20H00337) and Waseda University Grant for Special Research Projects (Project Nos. 2019C-253 and 2020C-269). KK is supported by World-leading Innovative Graduate Study Program for Materials Research, Industry, and Technology (MERIT-WINGS) of the University of Tokyo.

\end{document}